\documentclass[10pt]{IEEEtran}
\usepackage{enumitem}
\usepackage{etoolbox}
\usepackage{amsmath, amssymb, amsfonts, bbm, bm}
\usepackage[utf8]{inputenc}
\usepackage{longtable}
\usepackage{subfig}
\usepackage{flafter}
\usepackage{multirow}
\usepackage{pbox}
\usepackage{needspace}
\usepackage{cancel}
\usepackage{url}
\usepackage{rotating}
\usepackage{array,ragged2e}
\newcolumntype{P}[1]{>{\RaggedRight\arraybackslash}p{#1}}
\usepackage{footnote}
\usepackage[english]{babel}
\usepackage{graphicx}
\usepackage{lettrine}
\usepackage{subfig}
\usepackage{pifont}
\usepackage{authblk}
\usepackage{algpseudocode}

\newcommand{\abs}[1]{\lvert #1 \rvert}
\newcommand{\rtrunc}{\rfloor\!\vert}
\newcommand{\xmark}{\ding{55}}%
\usepackage[dvipsnames]{xcolor}

   \author{Ismael Martinez\thanks{I. Martinez is with the Department of Computer Science and Operations
Research, University of Montreal, Quebec, Canada H3C 3J7 (e-mail:
ismael.martinez@umontreal.ca).}, \and Abdelhakim Senhaji Hafid\thanks{A. S. Hafid is with the Department of Computer Science and Operations
Research, University of Montreal, Quebec, Canada H3C 3J7 (e-mail:
ahafid@iro.umontreal.ca).},
\and Michel Gendreau\thanks{M. Gendreau is with the Department of Mathematics and Industrial Engineering, Polytechnique Montreal, Quebec, H3C 3A7, Canada (email:michel.gendreau@polymtl.ca)}
}

\providecommand{\keywords}[1]{\textbf{\textit{Index terms---}} #1}

\title{FISIE: A Blockchain-Based Framework for Audit-Driven Integrity Enforcement in IoT--Fog Environments}
\begin{document}

\maketitle

\begin{abstract}
The success of smart city applications depends on secure and reliable interactions between IoT devices and edge computing infrastructures. Resource-constrained IoT devices often rely on fog nodes at the network edge to offload computation. However, fog nodes are not guaranteed to process requests correctly, whether due to faults, selfish behaviour, or malicious intent. Ensuring service integrity in decentralized IoT--fog environments remains a critical challenge. In this paper, we propose FISIE (\textit{Fog Identity \& Service Integrity Enforcement}), a blockchain-based framework that enables mutual authentication, streamlines service transactions, and enforces fog integrity. The framework introduces audit-driven rewards and penalties that modify reputation and collateral, influencing service traffic and profit. We model system behaviour using a latent-state dynamic graphical model and evaluate performance through discrete-event simulation across reputation, net profit, and node survival. Results show that while behavioural strategies influence short-term performance, long-term survival are governed by audit-driven enforcement, with only sustained honest behaviour ensuring stable participation and profitability.
\end{abstract}

\keywords Internet of Things, fog computing, blockchain, smart contracts, service integrity, service auditing, reputation management, integrity enforcement

\section{Introduction}
\lettrine{T}{he} \textit{Internet of Things} (IoT) is an ever-growing paradigm of interconnected sensors and computing devices. It has been widely adopted across public and private sectors, enabling automation and smart environments~\cite{kumar2019internet}. At scale, collaborative IoT networks form the foundation of smart city systems~\cite{zhang2020design}.

Due to limited computational resources, many IoT applications rely on external processing and storage. Cloud computing provides extensive resources, but network congestion and the large distance to data centres introduce high latency~\cite{ciscofog2015}, which is prohibitive for real-time applications such as healthcare~\cite{santos2018analyzing}, autonomous vehicles~\cite{yu2019deployment}, and multimedia streaming~\cite{do2015proximal}.

\textit{Fog computing} extends cloud services to the network edge, where geographically distributed micro-data centres, or fog nodes, provide low-latency support for IoT applications~\cite{bonomi2012fog}. Together, cloud and fog computing provide the computational backbone that supports large-scale IoT-driven smart city architectures~\cite{baccarelli2017fog, giordano2016smart, martinez2020design}.

Despite its growing importance, existing work on fog computing primarily focuses on infrastructure design~\cite{martinez2020design, martinez2024design} and resource allocation~\cite{jamil2022resource}, often overlooking the distinct objectives of IoT devices and fog nodes. IoT devices require secure, reliable, and real-time services, whereas fog nodes seek compensation and consistent service demand. Most importantly, current models implicitly assume that fog nodes behave with integrity. In practice, however, fog nodes may return faulty or malicious results~\cite{li2020auditing}, and existing systems provide no mechanism to verify correctness. This lack of accountability leaves service integrity dependent on trust, allowing unreliable fog nodes to return incorrect or incomplete results without being detected.

Blockchain has been explored as a solution for decentralized security, authentication, and access control in IoT systems~\cite{khan2018iot}. However, blockchain alone does not verify the correctness of fog computation. Service auditing provides a complementary mechanism for validating computation results and detecting faulty or malicious behaviour.

To address these challenges, we propose FISIE (\textit{Fog Identity \& Service Integrity Enforcement}), a blockchain-based framework that enables mutual authentication, streamlines service transactions, and enforces service integrity. The framework introduces audit-driven rewards and penalties that modify fog \textit{reputation} and \textit{collateral}, which influence service demand and profitability.

Our contributions are as follows:
\begin{itemize}
\item We design a smart contract that jointly supports \textbf{identity management}, \textbf{payment management}, and \textbf{audit-driven integrity enforcement} for IoT--fog interactions.
\item We formulate a \textbf{latent-state dynamic graphical model} describing how fog-node behaviour influences service traffic and profit across successive audit cycles.
\item We evaluate the proposed framework via \textbf{discrete-event simulation}, examining the impact of behavioural strategies on reputation, net profit, and node survival.
\end{itemize}

To the best of our knowledge, this is the first framework that enforces fog service integrity through smart contract-enabled, audit-driven enforcement. FISIE is implemented as a proof-of-concept on Ethereum~\cite{buterin2013ethereum}, while remaining compatible with other smart contract-enabled blockchain platforms.

The remainder of this paper is structured as follows. Section~\ref{section-rw} reviews related work. Section~\ref{section-design} defines the system design of FISIE. Section~\ref{section--penalty-enforcement} outlines the audit-driven incentive model.
Section~\ref{section-iimsc-state} provides an overview of on-chain state management via lookup tables. Sections~\ref{section-iimsc-identity}--\ref{section-iimsc-integrity} describe the identity management, payment management, and audit-driven integrity enforcement modules. Section~\ref{sec-pgm} presents the latent-state dynamic graphical model.  Section~\ref{sec-sim} presents the simulation-based evaluation. Section~\ref{section-consideration} discusses extensions and design considerations. Finally, Section~\ref{sec-conclusion} concludes the paper. 

\section{Related Work} \label{section-rw}

This work builds upon contributions from four complementary research areas. We review prior work in (i) IoT–fog authentication, (ii) blockchain-based monetization, (iii) fog data auditing, and (iv) blockchain–fog integration.

\subsection{IoT--Fog Authentication} 
A key element in securing any system is \textit{authentication}~\cite{khan2018iot}, which protects IoT and fog environments from attacks such as man-in-the-middle and replay attacks~\cite{almadhoun2018user}. 
In the blockchain-supported IoT--fog environment, we are interested in elliptic curve cryptography (ECC)-based \textit{mutual authentication}, wherein an IoT device and a fog node authenticate each other prior to communicating and data sharing~\cite{singh2021mutual}. 

Singh and Chaurasiya~\cite{singh2021mutual} propose a lightweight ECC-based mutual authentication scheme that relies on a centralized cloud data centre for key management. While computationally efficient, it introduces a single point of failure and requires full trust in the cloud.

Patwary et al.~\cite{patwary2020fogauthchain} propose a blockchain-based mutual authentication scheme using IoT and fog location data. However, it only supports stationary devices due to its reliance on static location for validation.

\subsection{Blockchain-Based Monetization}

Blockchain \textit{smart contracts} have previously been considered to streamline service transactions from IoT to fog~\cite{debe2020monetization, huang2018bitcoin}. Debe et al.~\cite{debe2020monetization} consider a monetization smart contract in which IoT devices deposit funds, which are then transferred by the smart contract to pay for fog services. 

Huang et al.~\cite{huang2018bitcoin} propose a smart contract where IoT devices lock a payment deposit prior to service execution. After receiving the service result, the IoT device may either accept the result and release payment to the fog node, or dispute the transaction to reclaim the deposit. However, each action invokes a separate smart contract call, which may incur substantial blockchain fees and overhead.

\subsection{Data Auditing for Fog Computing}

Several works propose auditing schemes to verify cached data at edge servers~\cite{li2020auditing, qiao2022lightweight, ding2022edge}. A vendor requests a hash of the stored data and compares it with its own to verify integrity.

Tian et al.~\cite{tian2019privacy} propose tagging IoT data as it moves through fog and cloud layers, enabling third-party auditors to verify integrity via zero-knowledge proofs. However, these approaches assume cooperative fog nodes. In a trustless setting, nodes may refuse to provide verification data.

If corruption is detected, fog nodes may receive corrected data~\cite{tian2019privacy}, assuming errors are accidental. 
Moreover, existing work does not impose penalties for malicious behaviour, allowing fog nodes to tamper with data without consequence. In public systems, zero-knowledge proofs of integrity must be requested, but servers lack incentives to comply~\cite{li2020auditing, qiao2022lightweight, ding2022edge}. 
Finally, \textit{service auditing}---verifying the correctness of computation rather than stored data---remains largely unexplored.

\subsection{Blockchain--Fog Integration}

Fog nodes and blockchains are both distributed, making their integration a natural approach to reducing communication latency. Blockchain-enabled fog nodes can host a local copy of the blockchain, reducing dependence on remote infrastructure.

Almadhoun et al.~\cite{almadhoun2018user} use blockchain-enabled fog nodes for IoT authentication, where performance depends on communication delays between IoT, fog, and blockchain. Resource requirements can be reduced using `light nodes', which store summarized blockchain data~\cite{liu2021blockchain, palai2018empowering}.

\begin{table*}[t]
    \caption{Sources of inspiration for FISIE.}
    \label{tab:inspiration}
    \centering
      \setlength{\tabcolsep}{6pt} % Default value: 6pt
    \renewcommand{\arraystretch}{1.15} % Default value: 1
    \begin{tabular}{>{\raggedleft\arraybackslash}P{0.103\linewidth}| P{0.095\linewidth}|P{0.08\linewidth}|P{0.107\linewidth}|P{0.08\linewidth}|P{0.08\linewidth}|P{0.08\linewidth} | P{0.07\linewidth}|P{0.09 \linewidth}}
       \textbf{ \newline $\phantom{.}$ \newline Contribution} & \textbf{Mutual \newline Authentication} & \textbf{Fog Service \newline Monetization} & \textbf{Resource Constrained \newline IoT-Compatible} & \textbf{Immutable \newline (Blockchain)} &
         \textbf{Fog \newline Penalization} & 
         \textbf{Fog Honesty \newline Verification} & \textbf{Public Fog \newline Auditing} & \textbf{Penalization for Malicious Fog} \\
         \hline \hline       Debe~\cite{debe2020monetization} & \xmark & \checkmark & \checkmark & \checkmark & \xmark & \xmark & \xmark & \xmark \\
         Almadhoun~\cite{almadhoun2018user} & \checkmark & \xmark & \checkmark & \checkmark & \xmark & \xmark & \xmark & \xmark \\
         Singh~\cite{singh2021mutual} & \checkmark & \xmark &  \checkmark & \xmark & \xmark & \xmark & \xmark & \xmark \\     
        Patwary~\cite{patwary2020fogauthchain} & \checkmark & \xmark &  \checkmark & \checkmark & \xmark & \xmark & \xmark & \xmark \\   
         Tian~\cite{tian2019privacy} & \xmark & \xmark & \xmark & \xmark & \xmark & \checkmark & \checkmark & \xmark \\
         Huang~\cite{huang2018bitcoin}  & \xmark & \checkmark & \xmark & \checkmark & \xmark & \checkmark & \xmark & \xmark \\
         \hline
         FISIE & \checkmark & \checkmark & \checkmark & \checkmark & \checkmark & \checkmark & \checkmark & \checkmark \\
         \hline \hline
    \end{tabular}
\end{table*}

\medskip

The gaps identified in Table~\ref{tab:inspiration} motivate the design of FISIE. We adopt standardized ECC parameters to remove reliance on centralized cloud servers for key management. We use blockchain-based lookup tables~\cite{patwary2020fogauthchain, cheng2021blockchain}, which facilitate both mutual authentication and streamlined service monetization. To enforce service integrity, FISIE introduces audit-driven incentives and deterrents~\cite{li2020auditing,qiao2022lightweight,ding2022edge}. Finally, FISIE assumes fog nodes that maintain only the subset of blockchain data required for authentication and auditing, i.e., \textit{integrity-focused light nodes}~\cite{liu2021blockchain,palai2018empowering}.

\section{FISIE System Design} \label{section-design}

We present the FISIE system design, which includes the physical architecture, cryptographic framework, IoT service request model, and smart contract modules.

\subsection{Physical Architecture}

The FISIE architecture is designed to be sufficiently general to be compatible with existing IoT--fog deployments.

\subsubsection{IoT Layer} 
The IoT layer consists of devices with heterogeneous resource capabilities. Let $I$ denote the set of IoT devices. Each device has sufficient resources to store cryptographic keys and communicate with the fog and cloud layers.

\subsubsection{Fog Layer} 
The fog layer consists of fog nodes, oracles~\cite{muhlberger2020foundational}, and blockchain nodes. Let $F$ denote the set of fog nodes and $O$ the set of oracles. Fog nodes have heterogeneous capabilities --- some may operate light blockchain nodes~\cite{palai2018empowering}, blockchain oracles~\cite{muhlberger2020foundational}, or both. Blockchain nodes may be deployed independently or integrated into fog nodes to form blockchain-enabled fog nodes~\cite{liu2021blockchain}.

\medskip

FISIE operates above existing IoT communication protocols~\cite{al2017internet}, providing integrity and incentive mechanisms without modifying the underlying communication infrastructure.

\subsection{Cryptographic Framework} \label{sec-ecc}

FISIE uses \textit{elliptic curve cryptography} (ECC) due to its strong security guarantees~\cite{kapoor2008elliptic} and native support in blockchain systems.
To ensure compatibility with existing blockchain infrastructure, we adopt Ethereum's ECC configuration, including cryptographic operations over the \texttt{secp256k1} curve~\cite{buterin2013ethereum,vujivcic2018blockchain,johnson2001elliptic}.

Each entity generates a key pair $(k, P)$, where the public key is derived from the private key as
\begin{align}
    P = k\cdot G_{\text{EC}},
\end{align}
for a fixed generator point $G_{\text{EC}}$ on the curve.

\paragraph{Key Agreement}
\textit{Elliptic Curve Diffie--Hellman} (ECDH) is used to establish a symmetric key for secure communication~\cite{haakegaard2015elliptic} between two entities $A$ and $B$ with key pairs $(k_A, P_A)$ and $(k_B, P_B)$:
\begin{align}
    k_A \cdot P_B = k_B \cdot P_A = k_A \cdot k_B \cdot G_{\text{EC}}.
\end{align}

\paragraph{Digital Signatures}
We use the \textit{Elliptic Curve Digital Signature Algorithm} (ECDSA), as implemented in Ethereum~\cite{buterin2013ethereum,johnson2001elliptic}. Given a message and its signature, the corresponding public key can be recovered, enabling identity verification without explicitly transmitting the public key.

\paragraph{Blockchain Addresses}
Let ${H: \{0,1\}^* \mapsto \{0,1\}^{256}}$ denote a cryptographic hash function. In Ethereum, this corresponds to \texttt{Keccak-256}~\cite{goldreich5foundations}. Blockchain addresses are derived from public keys as
\begin{align}
    a = H(P)\rtrunc_{20}, \label{eq-address}
\end{align}
where $\rtrunc_{20}$ denotes truncation to the right-most 20 bytes. 

\medskip

While Ethereum is used as the reference implementation, 
the design is blockchain-agnostic and can be deployed on alternative smart 
contract platforms such as Solana\footnote{https://solana.com/} or Layer~2 
solutions including Arbitrum\footnote{https://arbitrum.io/} and 
Optimism\footnote{https://www.optimism.io/} to improve scalability and reduce 
transaction costs~\cite{hafid2020scaling, thibault2022blockchain}.

\begin{figure}[t] 
\centering 
\includegraphics[width=0.8\linewidth]{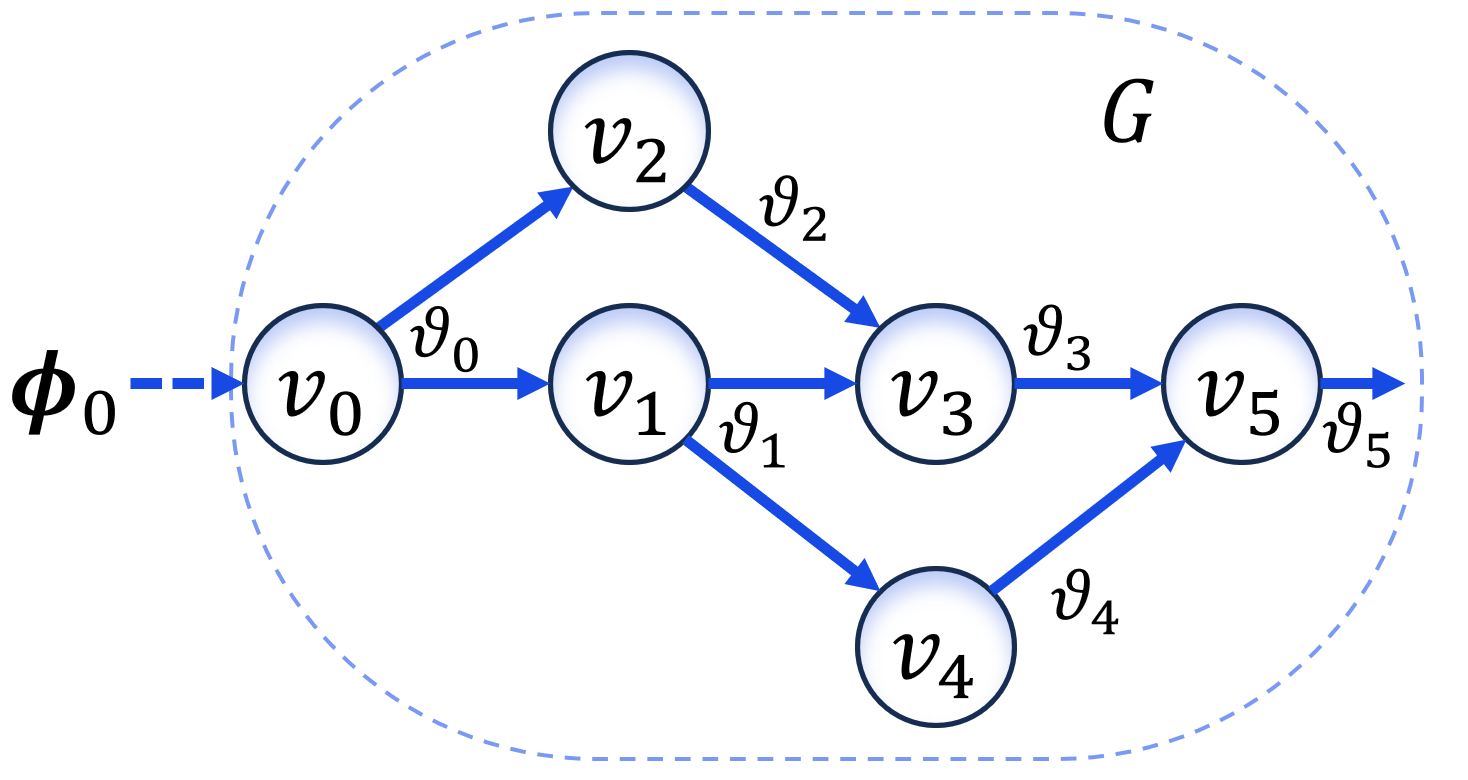} 
\caption{An IoT service request $\pi$ with task dependency graph $G$ and input vector $\boldsymbol{\phi}_0$.} 
\label{fig:tdg} 
\end{figure}

\subsection{IoT Service Requests as Task Dependency Graphs} \label{sec-tdg}

An IoT service request is modeled as a \textit{task dependency graph} (TDG)~\cite{liao2021dependency} structure $G = (\mathcal{V}, \mathcal{E})$, where the vertex set $\mathcal{V} = \{v_0,\dots,v_M\}$ represents computational tasks, and the edge set $\mathcal{E} \subseteq \mathcal{V}\times\mathcal{V}$ defines directed dependency relations between them. The vertices are indexed in a topological order, so that $(v_j,v_m)\in\mathcal{E}$ implies $j < m$~\cite{kahn1962topological}. The input to the initial task $v_0$ is the data vector $\boldsymbol{\phi}_0$. An IoT service request is therefore defined as $\pi = (G,\boldsymbol{\phi}_0)$.

Each task vertex $v_m \in \mathcal{V}$ represents a deterministic function that produces an output $\vartheta_m$. The task inputs and outputs are iteratively defined by
\begin{align}
        \boldsymbol{\phi}_m &= \begin{cases}
        \boldsymbol{\phi}_0 & \text{if }  m=0,  \\
            \{\vartheta_j \mid (v_j,v_m)\in\mathcal{E}\} & \text{if } m>0.
        \end{cases}
    \label{tdg-input} \\
    \vartheta_m &= v_m(\boldsymbol{\phi}_m), \quad v_m \in \mathcal{V}. \label{eq:tdg-output}
\end{align}
The evaluation of service request $\pi = (G,\boldsymbol{\phi}_0)$ is therefore \begin{align} \begin{split} \Pi(G,\boldsymbol{\phi}_0) &= \big( v_0(\boldsymbol{\phi}_0), \dots, v_M(\boldsymbol{\phi}_M)\big) \\ &= (\vartheta_0, \dots, \vartheta_M) = \boldsymbol{\vartheta}. \end{split} \label{eq-tdg-eval} \end{align}

This abstraction captures both sequential and parallel computation (Fig.~\ref{fig:tdg}). The task dependency structure enables selected intermediate outputs $\vartheta_m$ to be recomputed and verified without evaluating the full graph. This supports efficient spot-check auditing through \textit{partial verification}, described in Section~\ref{sec-iot-partial}.

For a request $\pi = (G,\boldsymbol{\phi}_0)$, let $c_\pi$ denote the processing cost, which depends on factors such as the size of $\boldsymbol{\phi}_0$ and the complexity of the task graph $G$. The request includes a proposed payment $\rho_\pi$ and may additionally specify a minimum reputation threshold $\tau_\pi$ that candidate fog nodes must satisfy.

\subsection{Smart Contract Modules}

The FISIE framework is enabled by the \textit{Identity \& Integrity Management Smart Contract} (IIMSC), which comprises three modules:

\paragraph{Identity Management}

This module governs entity registration and removal. These functions support IoT--fog mutual authentication.

\paragraph{Payment Management}

This module governs deposits, withdrawals, and IoT-to-fog payment allocation within the smart contract.

\paragraph{Audit-Driven Integrity Enforcement}

This module governs IoT-based reporting, reputation updates, and collateral penalties used to enforce service integrity on fog nodes.

\medskip

Detailed descriptions of these modules are provided in Sections~\ref{section-iimsc-identity}--\ref{section-iimsc-integrity}.

\section{Audit-Driven Incentive Model} 
\label{section--penalty-enforcement}

FISIE enforces fog integrity through service audits that influence \textit{reputation} and \textit{collateral}. Let $r_f(t) \in [0, R_{\max}]$ be the reputation score and $d_f(t) \in [0,D]$ be the remaining collateral of a fog node $f \in F$ at time $t$. Active fog nodes satisfy $r_f(t) \geq R_{\min}$ with $0 < R_{\min} < R_{\max}$ and $d_f(t) > 0$. Fog nodes violating these conditions are ejected from the system.

\subsection{Audit Coverage}

FISIE introduces two complementary audit mechanisms:
\mbox{(i) full} audits conducted by oracles covertly posing as IoT devices (Section~\ref{oracle-audit}), which preferentially target low-reputation fog nodes (Section~\ref{sec-oracle-selection}); and
(ii) partial verification conducted during IoT service transactions (Section~\ref{sec-iot-partial}), which preferentially interact with high-reputation fog nodes (Section~\ref{sec-iot-selection}).
Together, these complementary mechanisms provide verification coverage across the full reputation spectrum.

\subsection{Audit-Driven Incentives and Deterrents}

For a processed service request $\pi$, IoT partial verification and covert oracle audits create complementary monetary deterrents. A failed IoT partial verification may result in payment $\rho_\pi$ being withheld, whereas a failed covert oracle audit imposes a collateral penalty $d^-$. Together these mechanisms directly affect short-term earnings and create immediate \textbf{monetary incentives} and \textbf{monetary deterrents}.

Audit outcomes additionally modify the fog-node reputation score. Successful audits increase reputation by $r^+$, while failed audits decrease reputation by $r^-$. As a result of reputation-based selection, higher reputation increases IoT service demand, whereas lower reputation increases higher audit scrutiny. Consequently, reputation creates long-term \textbf{service incentives} and \textbf{service deterrents}.

Sustained audit failures compound these deterrents over time, resulting in ejection from the system when the reputation falls below the minimum threshold $(r_f(t) < R_{\min})$ or the collateral deposit is depleted ${(d_f(t) = 0)}$.

\section{IIMSC State Management} \label{section-iimsc-state}

The IIMSC smart contract maintains the persistent system state through blockchain-based lookup tables, internal fund allocation, and system parameters.

\subsection{IIMSC Lookup Tables} \label{sec-lookup}

The IIMSC smart contract maintains blockchain-based \textit{lookup tables} that store on-chain state information associated with registered IoT devices, fog nodes, and oracles. 

Since our default implementation uses Ethereum, all payments, collateral, and funds are denominated in Ether~\cite{buterin2013ethereum}.

\subsubsection{IoT Lookup Table $(T_I)$}
Each device $i \in I$ has a record $\mathbf{t}_i$ with the following fields:
\begin{itemize}
    \item \textbf{Address}: $\mathbf{t}_i.A = a_i$
        \item \textbf{PublicKey}: $\mathbf{t}_i.P = P_i$ 
        \item \textbf{AvailFunds}: $\mathbf{t}_i.AF$
\end{itemize}

\subsubsection{Fog Lookup Table $(T_F)$}

Each fog node $f \in F$ has a record $\mathbf{t}_f$ with the following fields:
\begin{itemize}
    \item \textbf{Address}: $\mathbf{t}_f.A = a_f$
    \item \textbf{AvailFunds}: $\mathbf{t}_f.AF$
    \item \textbf{Reputation}: $\mathbf{t}_f.R = r_f(t)$
    \item \textbf{Collateral}: $\mathbf{t}_f.D = d_f(t)$
\end{itemize}

\subsubsection{Oracle Lookup Table $(T_O)$}
An oracle $o \in O$ has a record $\mathbf{t}_o \in T_O$ with:
\begin{itemize}
    \item \textbf{Address}: $\mathbf{t}_o.A = a_o$
\end{itemize}

\subsubsection{Report Lookup Table $(T_\Sigma)$}
A report $\varsigma$ for request $\pi$ over IoT--fog pair $(i,f)$ has a record $\mathbf{t}_{\varsigma} \in T_\Sigma$ with:
\begin{itemize}
    \item \textbf{ReportID}: $\mathbf{t}_\varsigma.ID$
    \item \textbf{Addresses}: $\mathbf{t}_{\varsigma}.A = (a_i, a_f)$
    \item \textbf{Request}: $\mathbf{t}_\varsigma.RQ = \pi$
    \item \textbf{Payment}: $\mathbf{t}_\varsigma.PMT = \rho_\pi$
    \item \textbf{FogOutput}: $\mathbf{t}_\varsigma.FO = \boldsymbol{\vartheta}_f$
\end{itemize}

\medskip

Registration in the lookup tables is required for accountability and transparency.  
Reports serve as evidence of failed verification.

By default, lookup tables are implemented within the IIMSC smart contract. Alternatively, they may be maintained off-chain via a \textit{reverse oracle}~\cite{muhlberger2020foundational}. In this case, each record includes a ``LastUpdateHeader'' field storing the corresponding blockchain header, enabling verifiable synchronization with the on-chain state.

\subsection{IIMSC Reserves and Allocation}

All available funds and collateral are pooled within the IIMSC reserves. Service payments and collateral penalties are handled through internal fund allocation, where the corresponding lookup table entries are updated.

For example, allocation of payment $\rho$ from IoT device $i$ to fog node $f$ results in
\begin{align}
\begin{split}
\mathbf{t}_i.AF \gets& \mathbf{t}_i.AF - \rho \\
\mathbf{t}_f.AF \gets& \mathbf{t}_f.AF + \rho.
\end{split} \label{eq:allocation}
\end{align}
That is, funds are not transferred on-chain unless explicitly withdrawn. This design reduces transaction overhead while maintaining an on-chain record of available-fund balance updates.

\subsection{IIMSC Initialization}

The \texttt{initialization} function instantiates the lookup tables $\{T_I, T_F, T_O, T_\Sigma\}$ and sets the system parameters:
\begin{itemize}
    \item reputation parameters $\{R_{\text{init}}, R_{\min}, R_{\max}\}$
    \item reputation update parameters $r^+$ and $r^-$ with $r^- > r^+$
    \item initial collateral $D$ and penalty $d^-$
\end{itemize}

\section{Identity  Management} \label{section-iimsc-identity}

The Identity Management module of IIMSC governs entity registration and removal from the lookup tables, providing the verified identities required for IoT--fog mutual authentication.

\subsection{Registration}

Once IIMSC has been initialized, any entity that wishes to participate in FISIE must first register with the blockchain.
\subsubsection{IoT Registration}
The \texttt{iot\_registration} function takes initial Ether funds $E_q$ of amount $q>0$ and public key $P_i$ from IoT device $i\in I$. After verifying that the submitted public key $P_i$ corresponds to the caller address $a_i$, and ensuring address $a_i$ is not already in $T_I$,
the values $\mathbf{t}_i.A \gets a_i$, $\mathbf{t}_i.P \gets P_i$ and $\mathbf{t}_i.AF \gets q$ are added to new record $\mathbf{t}_i \in T_I$.

\subsubsection{Fog Registration}
The \texttt{fog\_registration} function takes initial Ether funds $E_q$ of amount $q$ from fog node $f \in F$, with ${q \geq D}$. The submitted amount $q$ is partitioned into: \mbox{(i) collateral $D$} , and (ii) available funds $q - D$. After ensuring address $a_f$ is not already in $T_F$, the values ${\mathbf{t}_f.A \gets a_f}$, $\mathbf{t}_f.D \gets D$, $\mathbf{t}_f.AF \gets q-D$ and initial reputation score $\mathbf{t}_f.R \gets R_{\mathrm{init}}$ are added to the new record $\mathbf{t}_f \in T_F$.

\subsubsection{Oracle Registration}
An oracle $o \in O$ must register with the blockchain in order to invoke IIMSC audit functions. We define a function $\texttt{oracle\_registration}$. After ensuring address $a_o$ is not already in $T_O$, the values $\mathbf{t}_o.A \gets a_o$ is added to the new record $\mathbf{t}_o \in T_O$.

\subsection{Removal}

An IoT device $i\in I$ that wishes to exit the system may call the \texttt{iot\_remove} function. All available funds $\mathbf{t}_i.AF$ are returned to IoT device $i$, and the record $\mathbf{t}_i$ is removed from $T_I$. 

A fog node may request to exit the system, or may be ejected by IIMSC if $\mathbf{t}_f.D = 0$ or $\mathbf{t}_f.R < \text{IIMSC}.R_{\min}$. In both cases, the \texttt{fog\_remove} function is executed. It returns any remaining collateral $\mathbf{t}_f.D$ and available funds $\mathbf{t}_f.AF$ fog node $f$, and the record $\mathbf{t}_f$ is removed from $T_F$. 

\subsection{Mutual Authentication} \label{sec-mutual-auth}

We outline the mutual authentication process between an IoT device $i \in I$ and a fog node $f \in F$. The protocol uses nonce-based challenge--response~\cite{koien2015brief}, where each party issues a unique nonce to prevent replay attacks. 

\medskip

\needspace{2\baselineskip}
\noindent \textbf{Protocol:}
\begin{enumerate} 
\item IoT device $i$
\begin{enumerate}[label=(\alph*)]
    \item requests nonce $n_f$ from fog node $f$.
    \item sends nonce $n_i$ to fog node $f$.
\end{enumerate}
    \item Fog node $f$ signs the message $\mu = n_i$ to produce signature $s_f$, and sends $(s_f,n_f)$ to IoT device $i$, where $n_f$ is the fog node's nonce.
    \item IoT device $i$ recovers $P_f$ from $s_f$ and $\mu = n_i$, and derives ${\tilde{a}_f = H(P_f)\rtrunc_{20}}$. It verifies that $\tilde{a}_f = \mathbf{t}_f.A$ for $\mathbf{t}_f \in T_F$. \label{m-a-query-f}
    \item If verification succeeds in step~\ref{m-a-query-f}, proceed. Otherwise, \textcolor{red}{mutual authentication fails}.
    \item IoT device $i$ signs the message $\mu = n_f$ to produce signature $s_i$, and sends $s_i$ to fog node $f$.
    \item Fog node $f$ recovers the public key $P_i$ from $s_i$ and ${\mu = n_f}$, and verifies that $P_i = \mathbf{t}_i.P$ for $\mathbf{t}_i \in T_I$. \label{m-a-query-i}
    \item If verification succeeds in step~\ref{m-a-query-i}, proceed. Otherwise, \textcolor{red}{mutual authentication fails}.
    \item IoT device $i$ and fog node $f$ establish a symmetric key via ECDH for secure communication:
\begin{align}
    k_i \cdot P_f = k_f \cdot P_i.
\end{align}
\label{m-a-connect}
\end{enumerate}

\section{Payment Management} \label{section-iimsc-payment}

The Payment Management module of IIMSC governs deposits, withdrawals, and payment allocation within the smart contract, supporting IoT--fog service transactions.

\subsection{Deposits and Withdrawals}

The following functions manage available funds.

\subsubsection{IoT Funds}
After an IoT device $i \in I$ has registered with the blockchain, it may deposit funds into its available funds reserve. The \texttt{iot\_deposit\_funds} function takes Ether funds $E_q$ of amount $q>0$ from IoT device $i$, updating ${\mathbf{t}_i.AF \gets \mathbf{t}_i.AF + q}$.

Similarly, \texttt{iot\_withdraw\_funds} takes a value ${q \in (0,\mathbf{t}_i.AF]}$, updating $\mathbf{t}_i.AF \gets \mathbf{t}_i.AF - q$, and transferring $q$ Ether to device $i$.

\subsubsection{Fog Funds}
Fog node $f \in F$ accumulates service payments within its available funds $\mathbf{t}_f.AF$. The \texttt{fog\_withdraw\_funds} function allows withdrawal of ${q \in (0,\mathbf{t}_f.AF]}$, updating ${\mathbf{t}_f.AF \gets \mathbf{t}_f.AF - q}$, and transferring $q$ Ether to $f$.

\subsection{IoT--Fog Service Transaction} \label{service-transaction}

For a request $\pi$, IoT device $i \in I$ first selects fog nodes satisfying a reputation threshold ${r_f(t) \geq \tau_\pi}$. After selecting a candidate node and completing mutual authentication, IoT device $i$ may request service from fog node $f \in F$.

\medskip
%\needspace{2\baselineskip}
\noindent \textbf{Protocol:}
\begin{enumerate}
    \item IoT device $i$ sends a service request $\pi$ and proposed payment $\rho_\pi$ to fog node $f$.
    \item Fog node $f$ either accepts or rejects the request. If rejected, either explicitly or by timeout, \textcolor{red}{the protocol terminates}.
    \item If accepted, fog node $f$ evaluates the request and returns the output vector $\boldsymbol{\vartheta}_f$.
    \item If IoT device $i$ accepts the output vector $\boldsymbol{\vartheta}_f$,
\begin{enumerate}[label=(\alph*)]
        \item IoT device $i$ invokes \texttt{iot\_fog\_transaction} with fog address $a_f$ and payment $\rho_\pi$.
        \item the function allocates $\rho_\pi$ from $\mathbf{t}_i.AF$ to $\mathbf{t}_f.AF$ as defined in~(\ref{eq:allocation}).
    \end{enumerate}
\end{enumerate}

The acceptance criteria for output vector
$\boldsymbol{\vartheta}_f$ are described in
Section~\ref{sec-iot-partial}.

\section{Audit-Driven Integrity Enforcement} \label{section-iimsc-integrity}

The Audit-Driven Integrity Enforcement module of IIMSC governs the reputation and collateral updates resulting from IoT verification and oracle audits.

\subsection{IoT-Based Partial Verification and Reporting} \label{sec-iot-partial}

IoT devices may perform \textit{partial verification} of fog computations. This provides lightweight integrity monitoring between oracle audits.

For a request $\pi=(G,\boldsymbol{\phi}_0)$,
IoT device $i$ receives the output vector
$\boldsymbol{\vartheta}_f$
returned by fog node $f\in F$, and selects a connected subgraph
$G'=(\mathcal V',\mathcal E') \subseteq G$ for verification. For each
$v_m\in\mathcal V'$,
the input $\boldsymbol{\phi}_m$ is generated according to~(\ref{tdg-input}),
where predecessors within
$\mathcal V'$
are recomputed and predecessors outside
$\mathcal V'$
are taken directly from
$\boldsymbol{\vartheta}_f$.

 The IoT device then computes
\begin{align}
 v_m(\boldsymbol{\phi}_m)=\vartheta_{i,m}, \quad \forall v_m \in \mathcal{V}'.
\end{align}

\paragraph{Positive Partial Result} 

If $\mathcal{V}'=\varnothing$ (trivial case) or
$\vartheta_{i,m}=\vartheta_{f,m}$ for all
$v_m\in\mathcal{V}'$, the verification succeeds.

The IoT device invokes \texttt{iot\_fog\_transaction},
which allocates payment as described in
Section~\ref{service-transaction} and applies a reputation
increase $r^+$:
\begin{align}
    \mathbf{t}_f.R \gets \min\{\mathbf{t}_f.R + r^+, R_{\max}\}.
\end{align}

\paragraph{Negative Partial Result}  
If there exists $v_m \in \mathcal{V}'$ such that $\vartheta_{i,m} \neq \vartheta_{f,m}$, 
IoT device $i$ withholds payment $\rho_\pi$ and invokes \texttt{iot\_report}, which takes the addresses $(a_i, a_f)$, the request $\pi$, payment $\rho_\pi$, and the output vector $\boldsymbol{\vartheta}_f$.  
This function creates a report $\varsigma$ and records it in the lookup table $T_\Sigma$,
which provides an auditable record of the failed verification.

In addition, the function applies a reputation decrease $r^-$:
\begin{align}
    \mathbf{t}_f.R \gets \max\{\mathbf{t}_f.R - r^-, 0\}.
\end{align}

\subsection{Ring Signatures for Anonymous IoT Set Membership} \label{sec-ring}

To prevent fog nodes from distinguishing covert oracle audits from normal IoT transactions, oracles operate through anonymous IoT identities. This requires proving membership in the IoT set without revealing which identity is being used.

Each oracle $o \in O$ maintains two distinct key pairs: an oracle key pair $(P_o, k_o)$ associated with blockchain address $a_o$ in $T_O$, and an IoT key pair $(P_\omega, k_\omega)$ associated with blockchain address $a_\omega$ in $T_I$. Since these identities are registered separately in distinct lookup tables, the IoT identity remains unlinkable to the oracle role, enabling covert audits.

This is achieved using a zero-knowledge proof of set membership, widely used in blockchain security~\cite{xu2021div}. 
We implement this proof using a ring signature scheme~\cite{chung2009ring}.

In the following, $x\|y$ denotes the concatenation of $x$ and $y$.

\medskip

\needspace{2\baselineskip}
\noindent \textbf{Setup:}
Let $P_I$ denote the set of public keys  in $T_I$:
\begin{align}
    P_I = \{\mathbf{t}_i.P \mid \mathbf{t}_i \in T_I\}.
\end{align}
We choose some enumerated subset $\bar{P} \subseteq P_I$ containing $P_\omega$ such that $\abs{\bar{P}} = L$. Let $P_\omega = P_j$ for some index $j \in \{1,\dots,L\}$.

An oracle with signing key pair $(P_j, k_j)$, corresponding to $P_\omega$, builds a ring signature of message $\mu$ over the elliptic curve defined in Section~\ref{sec-ecc}, 
with generator point $G_{\text{EC}}$ of prime order $p$. 

\begin{figure}[t]
    \centering
    \includegraphics[width=0.6\linewidth]{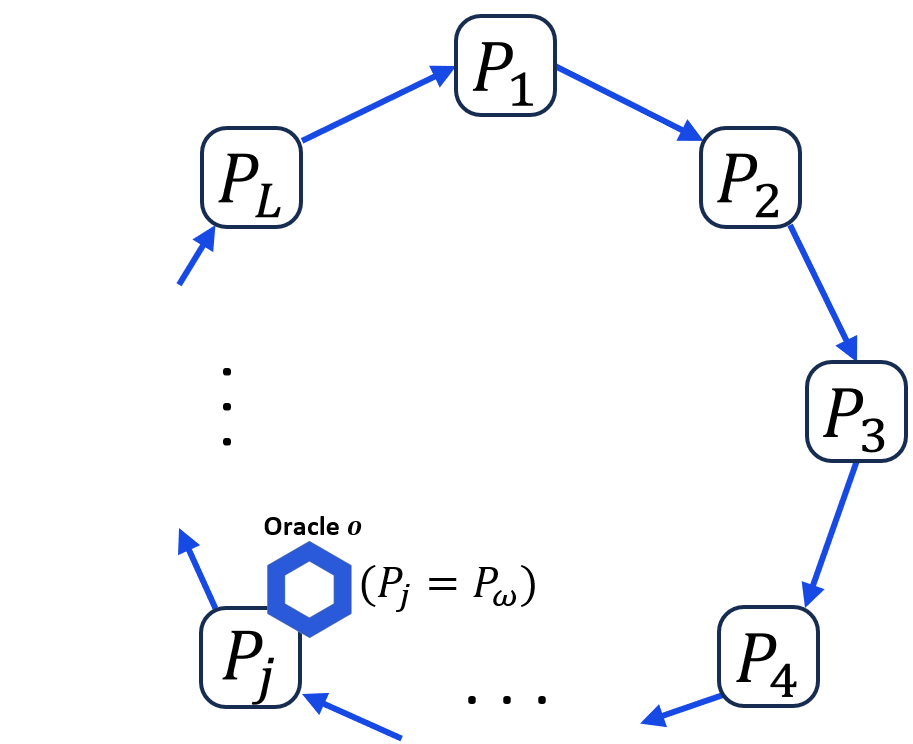}
    \caption{Ring signature over the IoT public-key set.}
    \label{fig:ring-sig}
\end{figure}

\needspace{2\baselineskip}
\noindent \textbf{Construction:}
\begin{enumerate}
    \item Choose a random integer $z \in [0, p-1]$. 
    \item Calculate $Q_j = (x_j, y_j) = z \cdot G_{\text{EC}}$.
    \item $\forall \ell=1,\dots, L$, $\ell\neq j$, randomly select integer ${\sigma_\ell \in [0, p-1]}$.
    \item  for $\ell = j+1, \dots, j+L-1 \ (\mathrm{mod}\ L)$:
    \begin{itemize}
        \item[] $e_\ell = H(\mu\|x_{\ell-1})$
        \item[] $Q_\ell = (x_\ell, y_\ell) = \sigma_\ell \cdot G_{\text{EC}} + e_\ell \cdot P_\ell$ 
    \end{itemize} 
    \item $e_j = H(\mu\|x_{j-1})$
    \item $ \sigma_j = z - e_j k_j$.
\end{enumerate}
The construction forms a cyclic chain of hash dependencies (‘ring’), ensuring no signer can be distinguished (Fig.~\ref{fig:ring-sig}).

Let $\boldsymbol{\sigma} = [\sigma_1, \dots, \sigma_L]$ and $\boldsymbol{P}=[P_1,\dots, P_L]$. 
The resulting ring signature is
\begin{align}
\mathcal{R}_\omega
=
(e_1,\boldsymbol{\sigma},\boldsymbol{P}). \label{eq-ring}
\end{align}

\medskip

\needspace{2\baselineskip}
\noindent \textbf{Verification:}

Given the ring signature $\mathcal{R}_\omega$~(\ref{eq-ring}), the verifier (IIMSC) begins with the provided value $e_1$ and computes
\begin{itemize}
    \item[] $Q_1 = (x_1, y_1) = \sigma_1 \cdot G_{\text{EC}} + e_1 \cdot P_1$
    \item[]  for $\ell=2,\dots, L$:
\begin{itemize}
        \item[] $e_\ell = H(\mu\|x_{\ell-1})$
        \item[] $Q_\ell = (x_\ell, y_\ell) = \sigma_\ell \cdot G_{\text{EC}} + e_\ell \cdot P_\ell$ 
    \end{itemize}
    \item[] $e'_1 = H(\mu\|x_L)$.
\end{itemize}
The ring signature is accepted if $e_1 = e'_1$.

\medskip

%\needspace{2\baselineskip}
\noindent \textbf{Correctness:}
By the original choice of $Q_j=z \cdot G_{\text{EC}}$, the final choice of $\sigma_j$ ensures correctness by `closing' the ring. That is,
\begin{align}
\begin{split}
Q_j   =& \ \sigma_j \cdot G_{\text{EC}} + e_j P_j \\
    =& \ (z - e_j  k_j) \cdot G_{\text{EC}} + e_j k_j \cdot G_{\text{EC}} \\
        =& \ z \cdot G_{\text{EC}}.
    \end{split}
\end{align}

\subsection{Covert Oracle Audits, Rewards and Penalties} \label{oracle-audit}

Covert oracle audits perform full recomputation of fog-node outputs. Oracle $o$, operating under IoT identity $\omega$, participates in the service transaction protocol as a normal IoT device. Consequently, it invokes \texttt{iot\_fog\_transaction} before performing the audit.

Oracle $o$ submits audit request ${\pi = (G, \boldsymbol{\phi}_0)}$ to fog node $f \in F$. Let $\boldsymbol{\vartheta} = \Pi(G,\boldsymbol{\phi}_0)$ denote the correct output vector defined by~(\ref{eq-tdg-eval}), and let $\boldsymbol{\vartheta}_f$ denote the output vector produced by fog node $f$. 
The audit outcome is represented by $X\in \{0,1\}$:
\begin{align}
    X = \begin{cases}
        0 & \text{if } \boldsymbol{\vartheta}_f \neq \boldsymbol{\vartheta} \\
        1 & \text{if } \boldsymbol{\vartheta}_f = \boldsymbol{\vartheta}.
    \end{cases}
\end{align}

The oracle records the audit result by invoking
\texttt{audit\_result},
which takes fog-node address $a_f$,
ring signature $\mathcal{R}_\omega$,
and audit outcome $X$.
The ring signature is generated over message
$\mu=a_f$.

\paragraph{Positive Audit Result ($X=1$)}

The oracle invokes
\texttt{audit\_result}$(a_f,\mathcal{R}_\omega,1)$.
No further action is required, since the reputation increase
applied through \texttt{iot\_fog\_transaction}
is consistent with the audit outcome.

\paragraph{Negative Audit Result ($X=0$)}

The oracle invokes
\texttt{audit\_result}$(a_f,\mathcal{R}_\omega,0)$.
Since the oracle previously invoked
\texttt{iot\_fog\_transaction}
while posing as an IoT device, the reputation increase applied during
the service transaction must be reversed. The smart contract therefore
updates
\begin{align}
\mathbf{t}_f.R
&\gets
\max\{\mathbf{t}_f.R-r^--r^+,0\},
\\
\mathbf{t}_f.D
&\gets
\max\{\mathbf{t}_f.D-d^-,0\}.
\end{align}

\paragraph{Collateral Redistribution} \label{sec-coll-redist}

The deducted collateral $d^-$ is redistributed among IoT devices to compensate for potential service disruption.

For each $i \in I$, available funds are updated as
\begin{align}
\mathbf{t}_i.AF \gets \mathbf{t}_i.AF + d_i,
\end{align}
such that
\begin{align}
\sum_{i \in I} d_i = d^-.
\end{align}

The distribution of collateral among IoT devices may be uniform or based on application-specific weighting schemes, such as request frequency.

\medskip 

\begin{figure}[t]
    \centering
    \subfloat[ECC-based mutual authentication (Section~\ref{sec-mutual-auth})]{\includegraphics[width=0.8\linewidth]{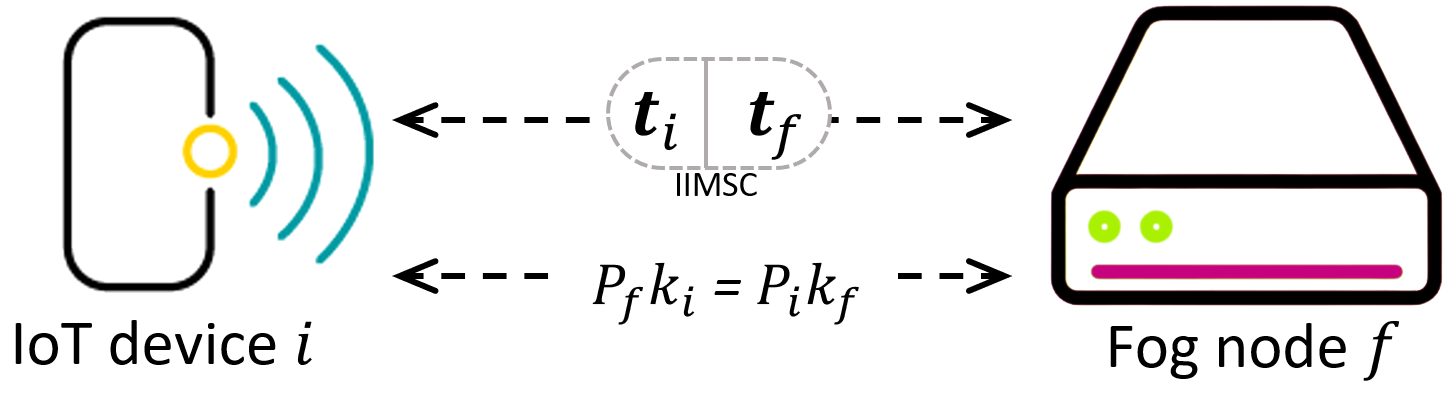}
    \label{subfig:mutual-authentication}}
    \hfill
    \subfloat[Streamlined service transaction (Section~\ref{service-transaction})]{\includegraphics[width=0.8\linewidth]{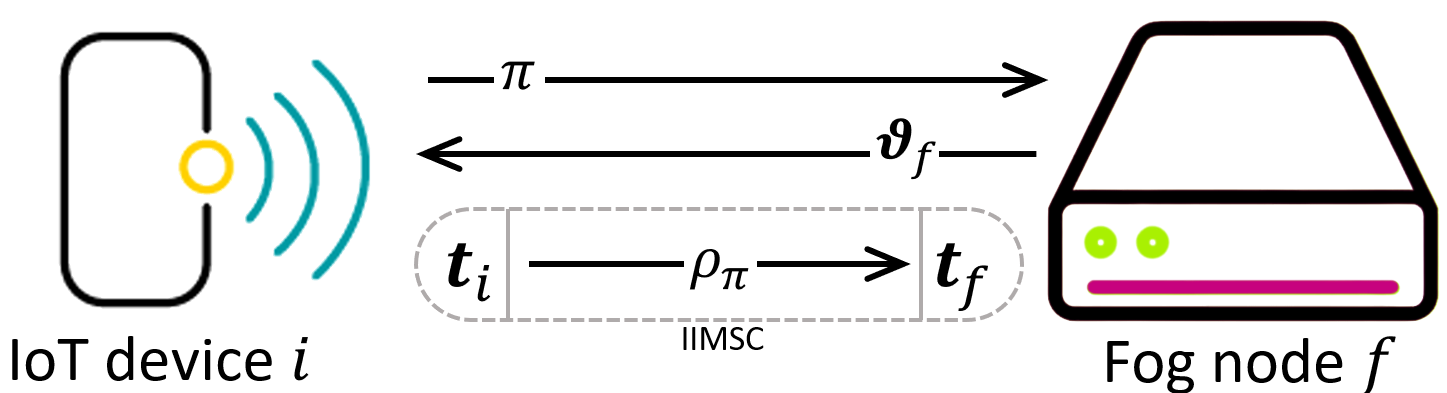}
    \label{subfig:service-transaction}}
    \hfill
    \subfloat[Audit-driven integrity enforcement (Section~\ref{oracle-audit})]{\includegraphics[width=0.8\linewidth]{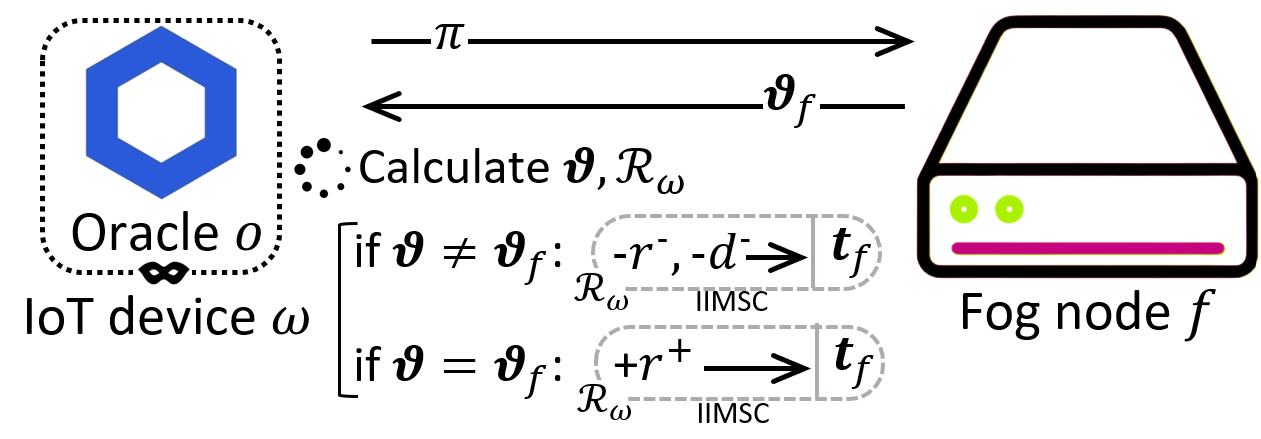}
    \label{subfig:integrity-enforcement}}
    \caption[]{Overview of the FISIE workflow.\footnotemark
(a) Mutual authentication establishes a symmetric session key.
(b) Service execution exchanges the request, output, and payment.
(c) Covert oracle audits verify computation correctness and enforce integrity through reputation and collateral updates.}
    \label{fig:fisie-overview}
\end{figure}

FISIE combines on-chain smart contract operations with off-chain IoT--fog and oracle--fog protocols. Table~\ref{tab:iimsc-functions} lists the IIMSC function interfaces, while Fig.~\ref{fig:fisie-overview} summarizes the workflows described in Sections~\ref{section-iimsc-identity}--\ref{section-iimsc-integrity}.

\begin{table}[!t]
    \centering
    \caption{IIMSC function interfaces.}
    \label{tab:iimsc-functions}
    \setlength{\tabcolsep}{6pt} % Default value: 6pt
    \renewcommand{\arraystretch}{1.1} % Default value: 1
    \begin{tabular}{p{0.16\linewidth}|p{0.73\linewidth}}
    \hline \hline 
    \textbf{Initialization} & \texttt{-initialization()}\\
    \hline
        \textbf{Registration} & \texttt{-iot\_registration(Ether $E_q$, \mbox{Public Key $P_i$})} \\
         & \texttt{-fog\_registration(Ether $E_q$)} \\
         & \texttt{-oracle\_registration()} \\
         \hline 
        \textbf{Removal} & \texttt{-iot\_remove()} \\
                & \texttt{-fog\_remove()} \\
                \hline
            \textbf{Funds} & \texttt{-iot\_deposit\_funds(Ether $E_q$)} \\
              & \texttt{-iot\_withdraw\_funds(float $q$)} \\
              &  \texttt{-fog\_withdraw\_funds(float $q$)} \\
              \hline
            \textbf{Service \mbox{Integrity}} & \texttt{-iot\_fog\_transaction(Address $a_f$, \mbox{float $\rho_\pi$})} \\
             & \texttt{-audit\_result(Address $a_f$, \mbox{Ring signature $\mathcal{R}_\omega$}, Result $X$)} \\
             & \texttt{-iot\_report}(\texttt{Address} $a_i$, \texttt{Address} $a_f$,  \texttt{Request} $\pi$, \texttt{float} $\rho_\pi$, \texttt{Output} $\boldsymbol{\vartheta}_f$)\\
            \hline \hline 
    \end{tabular}
\end{table}

\section{Latent-State Dynamic Graphical Model} \label{sec-pgm}

\footnotetext{IoT icons by \url{https://www.avsystem.com}}

We formulate a latent-state dynamic graphical model~\cite{murphy2002dynamic}
to describe how fog-node behaviour influences service traffic and profit across successive audit cycles.

\begin{figure}[t]
    \centering
    \includegraphics[width=0.75\linewidth]{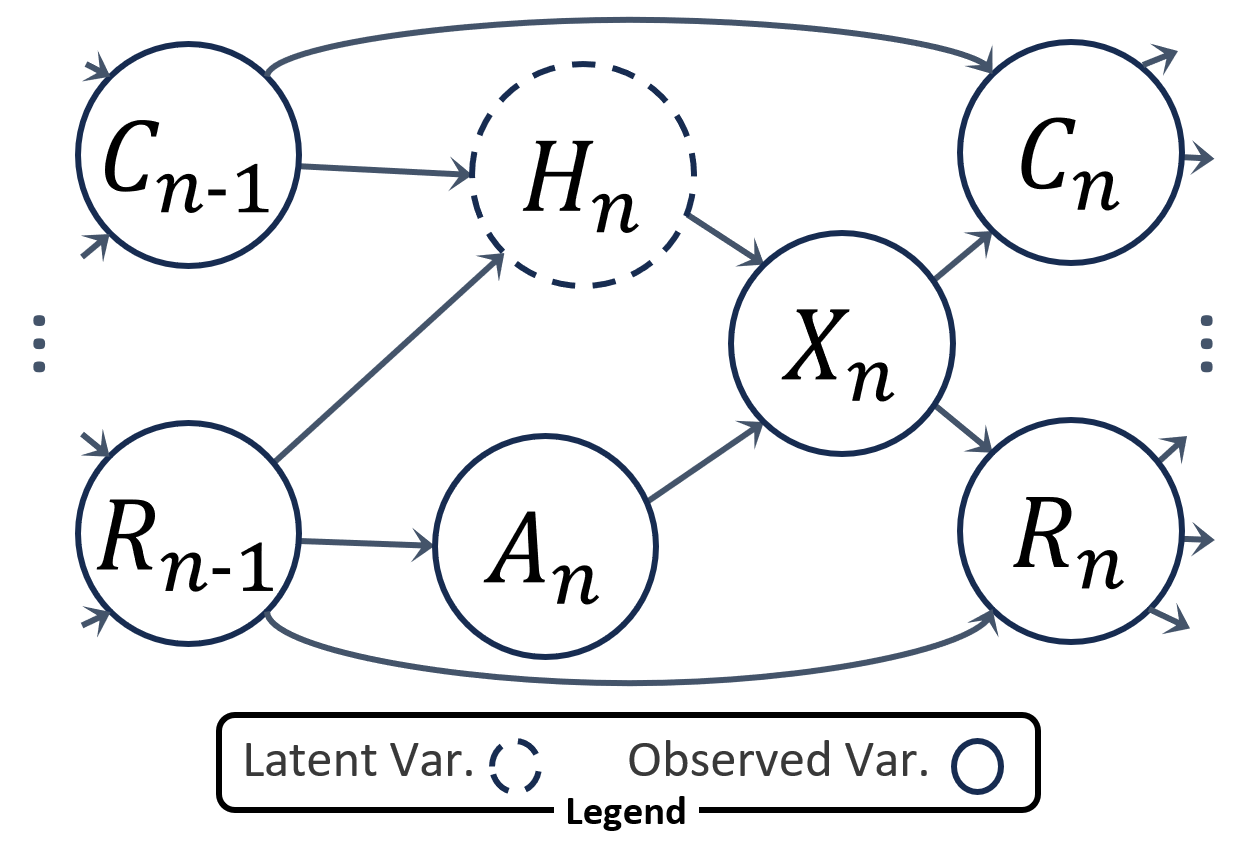}
    \caption{\textit{Latent-state dynamic graphical model} illustrating the dependencies between  audit-cycle variables.}
    \label{fig:pgm-workflow}
\end{figure}

Let $n$ denote the audit cycle index. As shown in Fig.~\ref{fig:pgm-workflow}, each audit cycle is characterized by the random variables
$A_n$, $H_n$, $X_n$, and $S_n$, where
\needspace{2\baselineskip}
\begin{itemize}
\item $A_n$ denotes the audit target selection outcome,
\item $H_n$ denotes the latent behavioural response of the audited fog node,
\item $X_n$ denotes the observed audit outcome,
\item $S_n=(R_n,C_n)$ denotes the reputation--collateral state of the audited fog node.
\end{itemize}

Each audit cycle consists of four stages: (i) audit target selection, (ii) fog-node behavioural response,
(iii) audit observation, and (iv) fog-node reputation--collateral state update. These stages correspond to the conditional distributions
\begin{align}
\text{Audit target selection:}\quad
& P(A_n \mid R_{n-1}), \\
\text{Behavioural response:}\quad
& P(H_n \mid S_{n-1}), \\
\text{Audit observation:}\quad
& P(X_n \mid H_n, A_n), \\
\text{State update:}\quad
& P(S_n \mid S_{n-1}, X_n).
\end{align}

For notational simplicity, time-dependent quantities such as $r_f(t)$ and $d_f(t)$ are written as $r_f$ and $d_f$ within this section.

\subsection{Audit Target Selection ($A_n$)} \label{behavioural-audit-selection}

The \textit{audit target selection} stage determines which fog node
$f \in F$ is selected by a covert oracle audit or an IoT service request. In the latent-state model, this corresponds to the conditional distribution $P(A_n \mid R_{n-1})$. Audit selection is conditioned on the observed reputation vector
\begin{align}
\mathbf{R} = \big(r_f\big)_{f \in F},
\end{align}
corresponding to the realization of $R_{n-1}$. 

\subsubsection{Oracle Audit Selection} \label{sec-oracle-selection}

Let $A_n = A_{o,f}^{\mathcal{O}}$ denote an oracle-initiated audit on fog node $f \in F$ by oracle $o \in O$. 
The audit target is selected according to a softmax distribution~\cite{franke2023softmax} applied to the negative reputation scores:
\begin{align}
P(A_{o,f}^{\mathcal{O}} \mid \mathbf{R})
&= \frac{\exp\!\left(-r_f\right)}
        {\sum_{f' \in F} \exp\!\left(-r_{f'}\right)}.
\label{eq:softmax-oracle}
\end{align}
Consequently, fog nodes with lower reputation are assigned higher audit probability.

\subsubsection{IoT Service Selection} \label{sec-iot-selection}

IoT devices may filter fog nodes based on reputation. Accordingly, each IoT device $i \in I$ imposes a minimum acceptable reputation threshold ${\tau_\pi \in [0, R_{\max}]}$ on a request $\pi$. Fog node $f$ is considered feasible if
\begin{align}
    r_f \geq \tau_\pi. \label{eq:thresh}
\end{align}

Let $\tilde{F}_\pi$ denote the set of feasible fog nodes:
\begin{align}
    \tilde{F}_\pi = \{ f \in F \mid r_f \ge \tau_\pi \}.
\end{align}

Let $\iota(\pi)$ map a request $\pi$ to its originating IoT device $i$, and let $A_n = A_{\iota(\pi),f}^{\mathcal{I}}$ denote IoT-initiated partial verification on fog node $f$ by IoT device $\iota(\pi)$. IoT service selection is performed according to a softmax distribution~\cite{franke2023softmax} over the reputation scores of feasible fog nodes:
\begin{align}
P(A_{\iota(\pi),f}^\mathcal{I} \mid \mathbf{R})
&= \frac{\exp\!\left(r_f\right)}
        {\sum_{f' \in \tilde{F}_\pi} \exp\!\left(r_{f'}\right)}.
\label{eq:softmax-iot}
\end{align} 
Consequently, fog nodes with higher reputation are favoured by IoT selection.

\medskip

Probabilistic selection prevents deterministic targeting and ensures that all fog nodes retain a non-zero probability of being audited. Consequently, malicious nodes cannot completely evade audits through  strategic behaviour.

\subsection{Behavioural Response ($H_n$)} \label{behavioural-honesty}

The \textit{behavioural response} stage governs how honestly a fog node $f$ responds to a request, given its current reputation $r_f$. In the latent-state model, this corresponds to the conditional distribution $P(H_n \mid S_{n-1})$.

Let $h_f(\pi) \in [0,1]$ denote the
\textit{intra-request honesty factor} applied to request $\pi$ by fog node $f$. This represents the fraction of a computational task performed correctly. For notational simplicity, we write $h_f$ within this section.

In general, the behavioural response differs between fog nodes and is not directly observable by the system. Therefore, we leave the conditional model $P(H_n \mid S_{n-1})$ unspecified at the latent-state level. Section~\ref{sec-behavioural} introduces several concrete behavioural honesty strategies and evaluate their long-term impact across successive audit cycles.

\subsection{Audit Observation ($X_n$)} \label{behavioural-observation}

The \textit{audit observation} stage corresponds to the conditional distribution $P(X_n \mid H_n, A_n)$ in the latent-state model. Let $H_n = h_f$ denote the latent honesty factor of the audited fog node $f$, and let $X_n \in \{0,1\}$ denote the observed audit outcome, where $X_n = 1$ indicates a pass and $X_n = 0$ a failure. The observed audit outcome determines both the reputation update and the monetary consequences experienced by the fog node.

\subsubsection{Oracle Audit Observation}
A covert oracle audit performs a full recomputation of the fog-node output and therefore accepts only fully correct computations. The observation of an oracle audit is defined as
\begin{align}
P(X_n = 1 \mid h_f, A_{o,f}^{\mathcal{O}})
&=
\begin{cases}
0, & h_f < 1, \\
1, & h_f = 1.
\end{cases}
\end{align}

\subsubsection{IoT Partial Verification Observation}

During a service transaction for request $\pi=(G,\boldsymbol{\phi}_0)$, an IoT device may perform
\textit{partial verification} (defined in Section~\ref{sec-iot-partial}) over a connected subgraph $G' \subseteq G$ before accepting the output vector returned by a fog node $f$.

Let $\nu_\pi \in [0,1]$ denote the \textit{verification fraction}, i.e., the fraction of request $\pi$ that is recomputed during partial verification. For analytical 
tractability, the subgraph selected for partial verification is represented as a contiguous \textit{verification segment} of normalized length $\nu_\pi$. Similarly, the connected subgraph $G'' \subseteq G$ that is computed correctly is represented as a contiguous \textit{honest segment} of normalized length $h_f$. 

Let $\eta_{\pi,f}$ denote the probability that a fog node with honesty factor $h_f$ passes partial verification with verification fraction $\nu_\pi$. Equivalently, $\eta_{\pi,f}$ is the probability that the sampled verification segment lies entirely within the honest segment of
the computation.

\medskip
\noindent\textbf{Trivial case:}
If no partial verification is performed $(\nu_\pi=0)$, or if the request is processed honestly in full $(h_f=1)$, then verification always passes. Therefore, $\eta_{\pi,f}=1$.

\medskip
\noindent\textbf{Impossible case:}
If $\nu_\pi>h_f$, the verification segment cannot be contained entirely
within the honest segment. Therefore, $\eta_{\pi,f}=0$.

\medskip
\noindent\textbf{Non-trivial case:}
Consider the case where $0<\nu_\pi\le h_f<1$.
Let $x \in [0,1-h_f]$ denote the left endpoint of the honest segment, and let $y \in [0,1-\nu_\pi]$ denote the left endpoint of the verification segment. The corresponding right endpoints are $x+h_f$ and $y+\nu_\pi$, respectively. 

\begin{figure}[t]
    \centering
    \subfloat[Partial verification pass]{
    \includegraphics[width=0.6\linewidth]{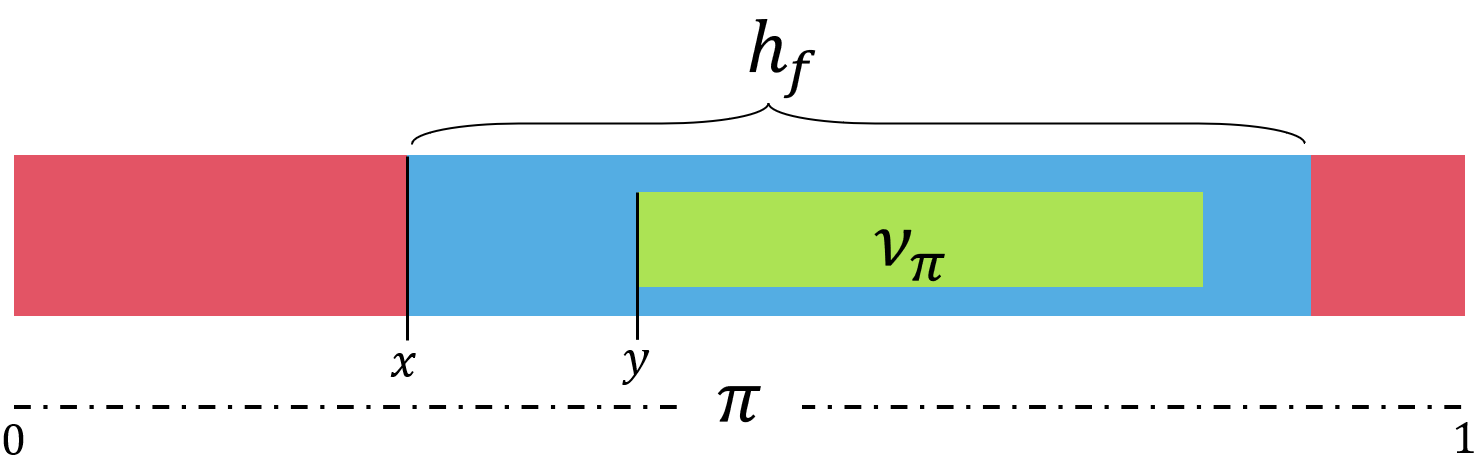}} \hfill
    \subfloat[Partial verification fail]{
    \includegraphics[width=0.6\linewidth]{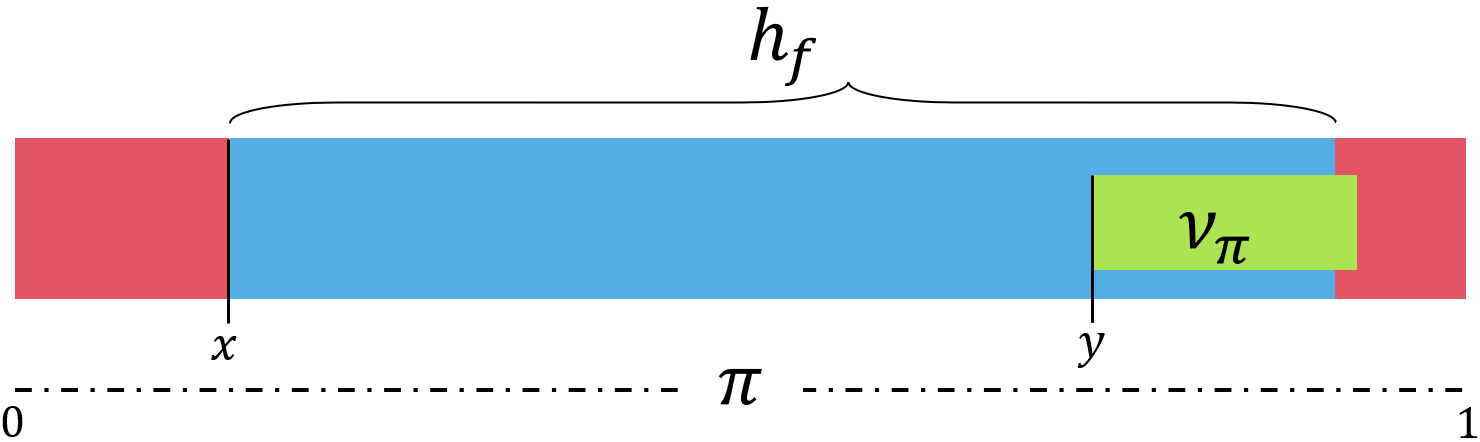}}
    \caption{A verification passes when the verification segment lies entirely within the honest segment of the computation. Otherwise, verification fails.}
    \label{fig:partial-intuition}
\end{figure}

Fig.~\ref{fig:partial-intuition} illustrates that a verification passes if and only if the verification segment lies entirely within the honest segment. That is, when $x \le y$ and $y+\nu_\pi \le x+h_f$. Merging these inequalities yields
\begin{align}
    x \le y \le x+h_f-\nu_\pi.
\end{align}

Since $x$ and $y$ are uniformly distributed over their respective domains, the probability of successful verification is
\begin{align}
\begin{split}
\eta_{\pi,f}
&=
\frac{1}{(1-h_f)(1-\nu_\pi)}
\int_{0}^{1-h_f}
\int_{x}^{x+h_f-\nu_\pi}
dy\,dx\\
&=
\frac{h_f-\nu_\pi}{1-\nu_\pi} < h_f.
\end{split}
\label{prob-pass}
\end{align}

\medskip
The resulting IoT partial verification observation is therefore defined as
\begin{align}
P(X_n = 1 \mid h_f, A_{\iota(\pi),f}^{\mathcal{I}}) &= \eta_{\pi,f},
\end{align}
where
\begin{align}
\eta_{\pi,f} =
\begin{cases}
0, & \nu_\pi > h_f, \\
\dfrac{h_f-\nu_\pi}{1-\nu_\pi}, & 0 < \nu_\pi \le h_f < 1, \\
1, & \nu_\pi = 0 \text{ or } h_f = 1.
\end{cases} \label{eq-pass-rate}
\end{align}

\subsection{Reputation--Collateral State Update ($S_n$)} \label{behavioural-reputation}

The \textit{reputation--collateral update} stage governs how the reputation and collateral of a fog node are updated following an audit. In the latent-state model, this corresponds to the conditional distribution ${P(S_n \mid S_{n-1}, X_n)}$, which in FISIE is deterministic. The corresponding state update is implemented by the audit-driven integrity enforcement module of the IIMSC smart contract.

\subsubsection{Reputation Update}
Recall that $r^+ > 0$ and $r^- > 0$ denote the reputation reward and penalty, respectively. We choose $r^- > r^+$ so that audit failures have a stronger impact than successes, discouraging dishonest behaviour.In the latent-state model, only the net reputation change is modelled. 

The audited fog node $f$ has a current reputation score ${r_f \in [R_{\min}, R_{\max}]}$. The reputation update is defined by
\begin{align}
r_f \leftarrow
\begin{cases}
\max\{r_f - r^-, 0\}, & \text{if } X_n = 0 \\
\min\{r_f + r^+,  R_{\max}\}, & \text{if } X_n = 1.
\end{cases}
\end{align}
If the update results in $r_f < R_{\min}$, fog node $f$ is ejected from the system.

\subsubsection{Collateral and Profit Update}
Recall that $d_f$ is the remaining collateral, and $d^-$ is the collateral penalty for an audit failure. 
Let $p_f$ denote the cumulative profit of fog node $f$ at time $t$, accounting for received service payments and computation costs. 
For a request $\pi$ and an honesty factor $h_f$, the fog node incurs reduced computation cost $h_f c_\pi$.

\paragraph{Covert Oracle Audit}
The remaining collateral is decreased by $d^-$ if the audit fails:
\begin{align}
d_f &\gets \max\{d_f - (1-X_n)d^-, 0\}.
\end{align}
If the update results in $d_f = 0$, fog node $f$ is ejected from the system.

To preserve oracle anonymity, payment $\rho_\pi$ is always released regardless of the audit outcome. Therefore,
\begin{align}
    p_f &\gets p_f + \rho_\pi - h_f c_\pi.
\end{align}

\paragraph{IoT partial verification}
Payment is released only if verification passes. Otherwise, payment is withheld.
\begin{align}
    p_f \gets p_f + X_n \rho_\pi - h_f c_\pi.
\end{align}
IoT partial verification does not change a fog node's collateral. 

\medskip 

The \textit{net profit} of fog node $f$, accounting for collateral $d_f$ locked in the system and initial collateral deposit $D$, is
\begin{align}
\psi_f = p_f + d_f - D.
\end{align}

\section{Simulation-Based Evaluation} \label{sec-sim}

We evaluate the proposed framework using discrete-event simulation~\cite{babulak2010discrete}. The complete simulation source code is publicly available on GitHub\footnote{\url{https://github.com/ismael-martinez/FISIE-DES}}.

We first define a set of behavioural honesty strategies to represent heterogeneous fog-node behaviour. We then specify the simulation parameters and analyze the resulting reputation, net profit, and node survival over successive audit cycles.

\subsection{Behavioural Honesty Strategies} \label{sec-behavioural}

We present several behavioural strategies corresponding to the conditional distribution
$P(H_n \mid S_{n-1})$ introduced in Section~\ref{behavioural-honesty}.
Consider a fog node $f \in F$, and request $\pi$ at time $t$. 
We model the honesty factor $h_f(\pi)$ as a Beta-distributed random variable.

Let $b_{\min} > 1$ and $b_{\max}$ denote the lower and upper bounds
used for behavioural strategies, i.e.,
$\alpha(t),\beta(t) \in [b_{\min}, b_{\max}]$. We define
\begin{align}
    h_f(\pi) \mid r_f(t), d_f(t)
    \sim \mathrm{Beta}\!\big(\alpha_f(t), \beta_f(t)\big).
\end{align}

The following strategies are either independent of fog-node reputation or adapt to it.
\subsubsection{Benchmark Strategies}
The following behavioural strategies are independent of fog-node reputation.

\paragraph{Conservative Strategy}
This strategy is biased toward high honesty.
It is modeled using a left-skewed distribution:
\begin{align}
    \alpha_f(t) = b_{\max}, \quad \beta_f(t) = b_{\min}.
\end{align}

\paragraph{Aggressive Strategy}
This strategy is biased toward low honesty.
It is modeled using a right-skewed distribution:
\begin{align}
    \alpha_f(t) = b_{\min}, \quad \beta_f(t) = b_{\max}.
\end{align}

\paragraph{Balanced Strategy}
This strategy is neutral with respect to honesty.
It is modeled using a symmetric distribution centred at moderate honesty.
\begin{align}
    \alpha_f(t) = \beta_f(t) = \frac{b_{\min} + b_{\max}}{2}.
\end{align}

\subsubsection{Adaptive Strategies}
The following behavioural strategies adapt expected honesty through an inverse relationship with either reputation or remaining collateral.

\paragraph{Opportunistic Strategy}
This strategy linearly adjusts honesty according to reputation. It is modeled as
\begin{align}
    \begin{split}
        \alpha_f(t) &= b_{\max} - (b_{\max}-b_{\min}) \frac{r_f(t) - R_{\min}}{R_{\max} - R_{\min}}, \\
        \beta_f(t)  &= b_{\min} + (b_{\max}-b_{\min}) \frac{r_f(t) - R_{\min}}{R_{\max} - R_{\min}}.
    \end{split}
\end{align}
High reputation allows the strategy to lower honesty and reduce computation costs under lower expected audit scrutiny. Conversely, when reputation is low, the strategy increases honesty to rebuild reputation.

\paragraph{Progressive Strategy}
This strategy linearly adjusts honesty according to remaining collateral. It is modeled as
\begin{align}
    \begin{split}
        \alpha_f(t) &= b_{\min} + (b_{\max}-b_{\min}) \frac{D-d_f(t)}{D - d^-}, \\
        \beta_f(t)  &=b_{\max} - (b_{\max}-b_{\min}) \frac{D-d_f(t)}{D - d^-}.
    \end{split}
\end{align}
High remaining collateral allows the strategy to reduce honesty and lower computation costs under a larger collateral buffer. Conversely, as collateral is depleted, the strategy progressively increases honesty to reduce the risk of ejection.

\subsection{Simulation Parameters}

The following parameters are used throughout the simulation study.

\subsubsection{System Parameters}

For the IIMSC smart contract parameters, we set $(R_{\min}, R_{\max}) = (2, 10)$, ${(r^+, r^-) = (1, 2.5)}$, and $(D, d^-) = (3, 1)$. 
Covert oracle audits and IoT service requests are modeled as independent Poisson processes~\cite{kingman1992poisson}. The IoT request rate is fixed at $\lambda^\mathcal{I}=40$. 
Two simulation parameter configurations are considered, differing only in the initial reputation $R_\mathrm{init}$ and oracle audit rate $\lambda^\mathcal{O}$: ${(R_\mathrm{init}, \, \lambda^\mathcal{O})=(7, 1)}$ and $(R_\mathrm{init}, \, \lambda^\mathcal{O})=(5, 2)$. Oracle audits remain substantially less frequent than routine IoT service requests in both parameter configurations.

Each fog node $f \in F$ enters the system with initial reputation
$r_f(0)=R_{\mathrm{init}}$, collateral $d_f(0)=D$, and net profit
$\psi_f(0)=0$. If the collateral is fully depleted before the reputation falls below $R_{\min}$, the fog node is ejected. This implicitly creates a \mbox{$D/d^- = 3$-strike} rule, while allowing sufficient room to rebuild reputation between violations.

\subsubsection{Behavioural Parameters}

\begin{figure}[t]
    \centering
    \includegraphics[width=0.85\linewidth]{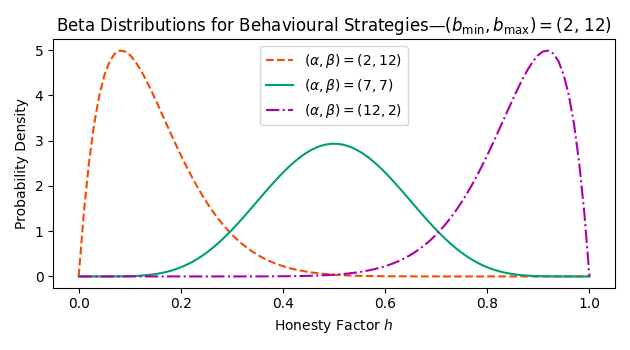}
    \caption{Beta distributions for behavioural strategies.}
    \label{fig:behavioural-beta}
\end{figure}

For the fog behavioural strategies of Section~\ref{sec-behavioural}, we set $b_{\min} = 2$ and $b_{\max} = 12$. These values yield strongly skewed Beta distributions under extreme parameterizations, with probability mass concentrated near $h_f(\pi)\approx 1$ or $h_f(\pi)\approx 0$, as illustrated in
Fig.~\ref{fig:behavioural-beta}. 

\subsubsection{Request Parameters}

For simulation, processing costs are modeled as
\begin{align}
    c_\pi \sim \mathrm{Gamma}(k=2,\theta=\tfrac14),
\end{align}
which captures non-negative, unbounded costs with most probability mass concentrated on smaller requests.

The proposed payment is modeled as a markup over cost:
\begin{align}
    \rho_\pi &= (1+\gamma_\pi)c_\pi,
\end{align}
where $\gamma_\pi \sim \mathrm{Beta}(2,10)$ is the profit margin.

Request reputation thresholds are modeled as
\begin{align}
    \tau_\pi \sim 10 \cdot \mathrm{Beta}(2,5).
\end{align}
Values $\tau_\pi < R_{\min}$ correspond to no additional filtering
beyond the system-wide minimum.

\subsubsection{IoT Verification Parameters}

The partial verification fraction $\nu_\pi$ is modeled as 
\begin{align}
\nu_\pi =
\begin{cases}
0, & \text{with probability } 0.2, \\
Z, & \text{with probability } 0.8,
\end{cases}
\end{align}
where $Z \sim \mathrm{Beta}(1,10)$. This concentrates probability mass near zero, reflecting the resource constraints of IoT devices and the preference for small partial verification fractions. Given $\nu_\pi$, the corresponding probability of passing partial verification $\eta_{\pi,f}$ is computed according to~(\ref{eq-pass-rate}).

\subsection{Simulation Setup \& Results} \label{sec-sim-results}

We evaluate FISIE via discrete-event simulation~\cite{babulak2010discrete}. We analyze outcomes in terms of reputation, net profit, and node survival. Each simulation is instantiated with 500 fog nodes per strategy, or 2500 nodes total. For each of the two parameter configurations, the simulation is repeated over 50 independent runs to reduce variance. Sample paths are aligned on a common time grid using last-value interpolation~\cite{kuffel1997implementation} and averaged pointwise.

\begin{figure*}[t]
    \centering
    \subfloat[Reputation averaged over active nodes, $(R_{\mathrm{init}},\,\lambda^{\mathcal O})=(7,1)$]{\includegraphics[width=0.475\linewidth]{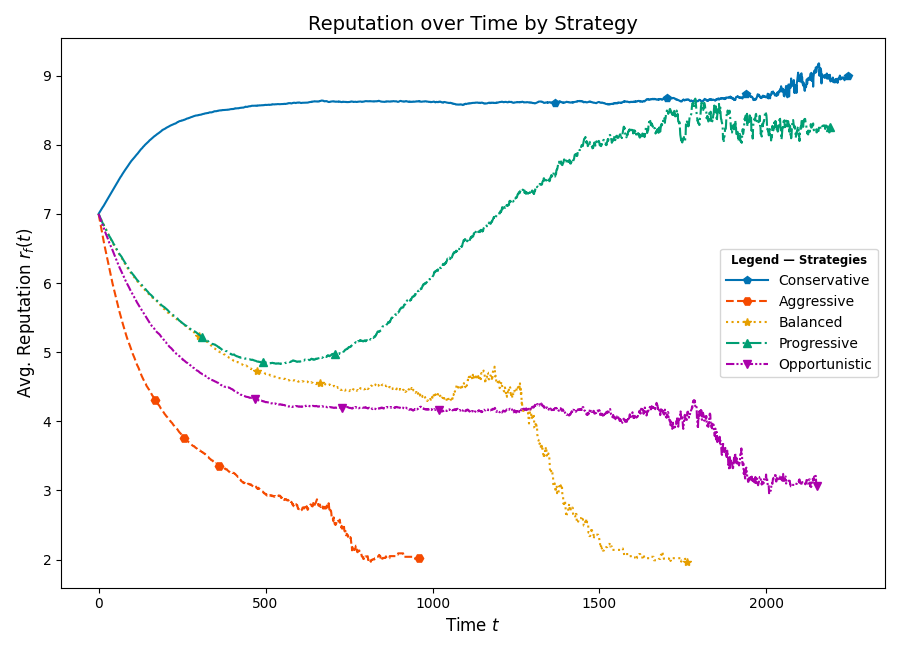}\label{subfig:rep025}} \qquad
        \subfloat[Reputation averaged over active nodes, $(R_{\mathrm{init}},\,\lambda^{\mathcal O})=(5,2)$]{\includegraphics[width=0.475\linewidth]{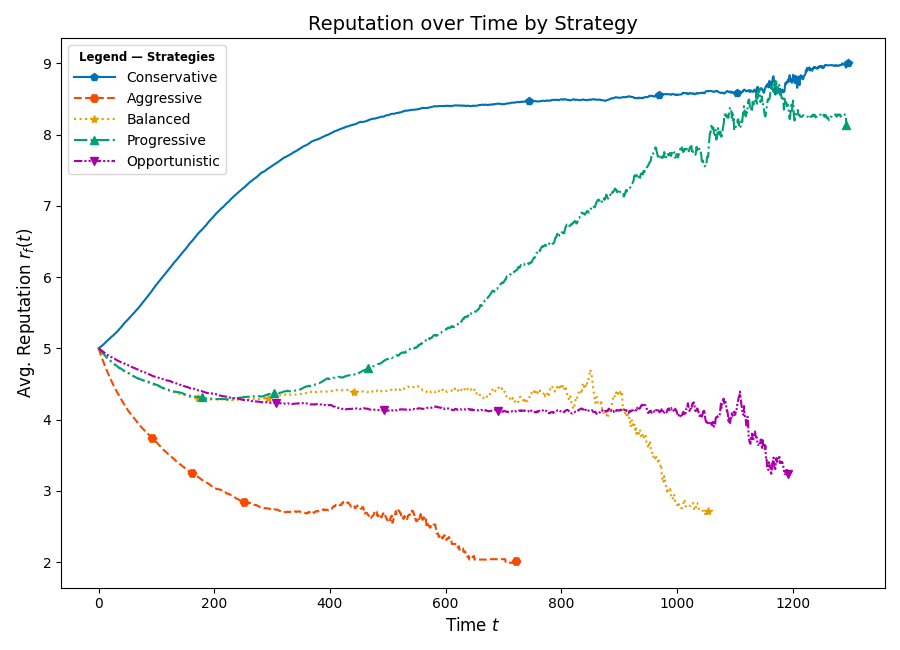}\label{subfig:rep05}}
    \hfill
    \subfloat[Net profit averaged over all nodes, $(R_{\mathrm{init}},\,\lambda^{\mathcal O})=(7,1)$]{\includegraphics[width=0.475\linewidth]{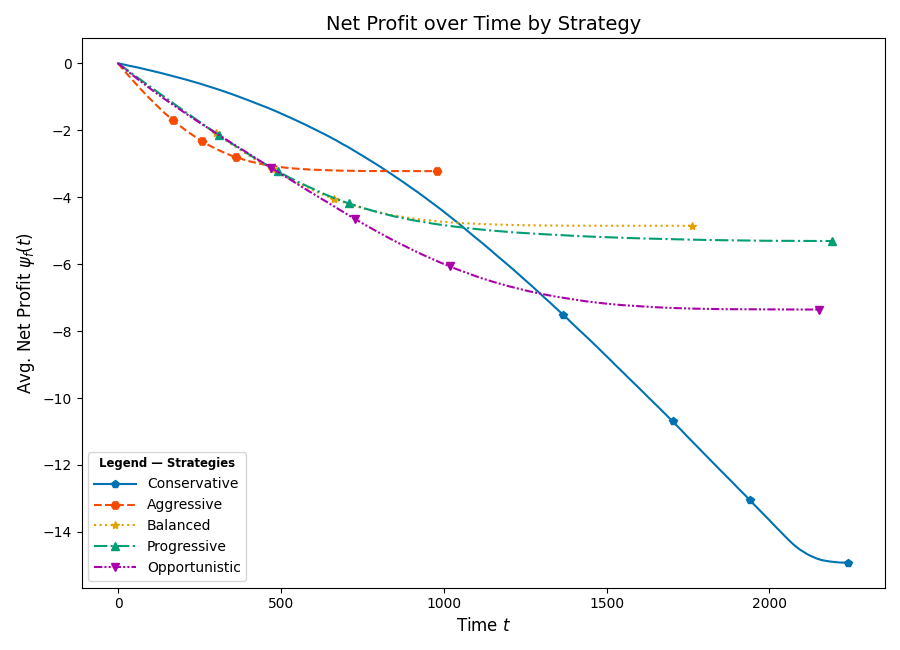}\label{subfig:profit025}}   \qquad
    \subfloat[Net profit averaged over all nodes, $(R_{\mathrm{init}},\,\lambda^{\mathcal O})=(5,2)$]{
        \includegraphics[width=0.475\linewidth]{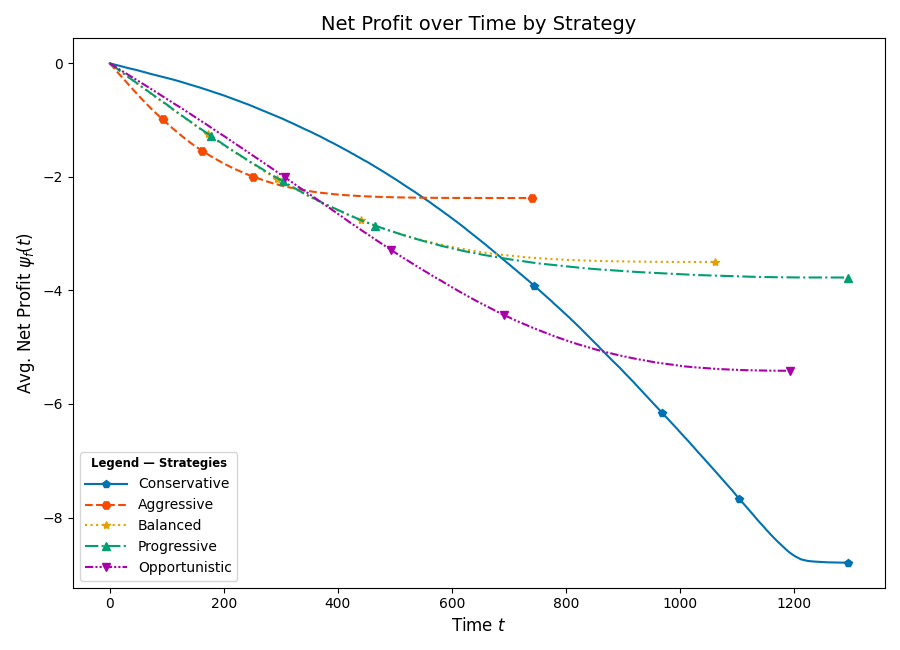}\label{subfig:profit05}}
        
    \caption{Reputation and net profit over successive audit cycles under two parameter configurations. Markers indicate the survival checkpoints at which 75\%, 50\%, 25\%, and 0\% of nodes remain active.}
    \label{fig:metric-strategy}
\end{figure*}

Results are presented in Fig.~\ref{fig:metric-strategy}, where node survival is indicated by checkpoint markers. Survival checkpoint statistics are reported in Table~\ref{tab:survival-checkpoints}. Checkpoint audit cycles count only verification events: covert oracle audits and positive IoT partial verification ($\nu_\pi>0$).

\begin{table}[t]
\caption{Survival checkpoint statistics by behavioural strategy under two simulation parameter configurations. Values correspond to the checkpoint markers in Fig.~\ref{fig:metric-strategy}.}
    \label{tab:survival-checkpoints}
    \centering
      \setlength{\tabcolsep}{6pt} % Default value: 6pt
    \renewcommand{\arraystretch}{1.1} % Default value: 1
    \subfloat[$(R_{\mathrm{init}},\,\lambda^{\mathcal O})=(7,1)$]{\begin{tabular}{r|P{0.15\linewidth}|P{0.15\linewidth}|P{0.15\linewidth}|P{0.15\linewidth}}
         & \multicolumn{4}{c}{\textbf{Checkpoint Audit Cycle\footnotemark (Mean $\pm$ SD)}}  \\
        \textbf{Strategy} & \textbf{75\%} & \textbf{50\%} & \textbf{25\%} & \textbf{0\%} \\
        \hline \hline
        Cons. & $43.7 \pm 1.5$ & $54.4 \pm 1.6$ & $62.1 \pm 1.6$ & $68.0 \pm 1.6$
 \\
        Aggr. & $5.4 \pm 0.2$ & $8.2 \pm 0.2$ & $11.6 \pm 0.3$ & $25.4 \pm 1.8$\\
        Bal. & $9.6 \pm 0.4$ & $15.2 \pm 0.5$ & $21.2 \pm 0.6$ & $44.9 \pm 3.3$\\
        Prog. &$9.9 \pm 0.5$ & $15.7 \pm 0.6$ & $22.6 \pm 1.0$ & $63.5 \pm 3.5$ \\
        Opp. & $15.0 \pm 0.7$ & $23.3 \pm 0.7$ &  $32.5 \pm 0.9$ & $60.6 \pm 3.2$ \\
        \hline \hline
    \end{tabular}} \hfill
    \subfloat[$(R_{\mathrm{init}},\,\lambda^{\mathcal O})=(5,2)$]{\begin{tabular}{r|P{0.15\linewidth}|P{0.15\linewidth}|P{0.15\linewidth}|P{0.15\linewidth}}
         & \multicolumn{4}{c}{\textbf{Checkpoint Audit Cycle\footnotemark (Mean $\pm$ SD)}}  \\
        \textbf{Strategy} & \textbf{75\%} & \textbf{50\%} & \textbf{25\%} & \textbf{0\%} \\
        \hline \hline
        Cons. & $23.8 \pm 1.1$ & $31.0 \pm 0.9$ & $35.3 \pm 0.9$ & $38.4 \pm 0.9$ \\
        Aggr. & $2.9 \pm 0.2$ & $5.2 \pm 0.2$ & $8.0 \pm 0.2$ & $19.2 \pm 1.7$ \\
        Bal. & $5.5 \pm 0.4$ & $9.3 \pm 0.4$ & $14.1 \pm 0.6$ & $30.6 \pm 1.8$\\
        Prog. &$5.6 \pm 0.4$ & $9.7 \pm 0.4$ & $14.9 \pm 0.7$ & $37.1 \pm 1.4$ \\
        Opp. & $9.8 \pm 0.4$ & $15.7 \pm 0.6$ & $22.1 \pm 0.7$ & $36.5 \pm 1.1$ \\
        \hline \hline
    \end{tabular}}
\end{table}
% \footnotetext{Reported in thousands of audit cycles.}

\subsubsection{Overall Results}

The behavioural trends remain consistent across both parameter configurations. Lower initial reputation and more frequent oracle audits shorten survival times for every strategy (Table~\ref{tab:survival-checkpoints}), while preserving the relative survival order: Conservative, Progressive, Opportunistic, Balanced, and Aggressive.

Across all strategies, average net profit remains negative because losses from failed IoT verification (withheld payments) and failed oracle audits (collateral penalties) outweigh the computation-cost savings obtained through reduced honesty. 
Consequently, strategies that retain a higher proportion of active nodes accumulate greater net losses over time. This trend is consistent across both parameter configurations.

\subsubsection{Benchmark Strategy Results}

For the Aggressive and Balanced strategies, ejection is primarily reputation-driven. Frequent IoT verification failures reduce reputation below the minimum threshold before collateral is depleted, causing Aggressive nodes to be ejected first, followed by Balanced nodes.
 
 In contrast, Conservative nodes rarely fail IoT verification and therefore maintain reputation above the minimum threshold throughout most of their lifetime. Their eventual removal is instead collateral-driven, resulting from the gradual depletion of collateral through less frequent oracle audits. Consequently, Conservative nodes remain active for substantially longer. As other nodes are ejected, the remaining active nodes receive an increasing share of oracle audits and IoT service requests, producing the larger cumulative losses observed for Conservative nodes in Fig.~\ref{subfig:profit025} and \ref{subfig:profit05}.

\footnotetext{Reported in thousands of audit cycles.}

 \subsubsection{Adaptive Strategy Results}

The Progressive strategy increases honesty after each collateral loss, allowing sruviving nodes to rebuild reputation over time. 
Although many Progressive nodes are initially ejected at rates similar to the Balanced strategy, surviving nodes progressively increase their honesty and rebuild reputation. Consequently, ejection of the remaining nodes shifts from being reputation-driven to collateral-driven, allowing them to survive nearly as long as Conservative nodes.

The Opportunistic strategy adjusts honesty according to reputation, causing nodes to regulate honesty according to their current reputation and maintain reputation near a stable operating region above the ejection threshold. Most eventual removals are therefore collateral-driven rather than reputation-driven.

\medskip

Fog-node behavioural strategies primarily affect short-term net profit, reputation, and survivability. Adaptive strategies can delay ejection by maintaining reputation above the minimum threshold for longer. However, nodes that avoid reputation-driven ejection remain subject to covert oracle audits, eventually leading to collateral-driven ejection.

These results demonstrate the effectiveness of the FISIE incentive structure. Periodic covert oracle audits eventually detect any sub-honest $(h<1)$ behaviour, causing penalties to accumulate until the node is removed through either reputation- or collateral-based ejection. Consequently, only fully honest $(h=1)$ behaviour sustains long-term participation and profitability.

\section{Extensions and Design Considerations} \label{section-consideration}

The preceding sections define the baseline design of FISIE. Several complementary mechanisms are omitted from this baseline, but can be incorporated to capture more complex operational settings.

\paragraph{Matching}  
IoT devices and fog nodes may participate in a mutual selection process prior to authentication. Matching can be based on proximity, resource availability, and other system-level criteria~\cite{tran2022survey}. Additionally, both parties may evaluate candidates based on \textit{reliability}, which incorporates reputation scores, collateral, and historical behaviour.

\paragraph{Bargaining}  
After authentication, an IoT device submits its request with a proposed payment. The fog node may counter-offer, initiating a bargaining process to determine a mutually agreed price~\cite{shih2019fog}. Alternatively, auction-based mechanisms can be used, where fog nodes competitively bid for service requests~\cite{peng2020multi}.

\paragraph{Disputes}

A fog node may dispute a report $\varsigma$ recorded in $T_\Sigma$ if it believes the claim is incorrect. The dispute can be escalated to an external decentralized resolution platform such as Kleros\footnote{https://kleros.io/}~\cite{metzger2018decentralized}. Since disputes incur costs, both parties are incentivized to act honestly.

If the dispute resolution process determines that the report was false, the IoT device is considered fraudulent. Possible countermeasures include temporary suspension of activity, reduced priority in future service requests, or exclusion from the system.

\paragraph{Verification Offloading}

Partial verification may be delegated to more capable entities, such as
oracles or nearby edge devices~\cite{samie2016computation}. This allows
resource-constrained IoT devices to benefit from stronger verification
without performing the computation locally.

\paragraph{Reputation and Collateral}

More sophisticated strategies for calculating~\cite{yu2010survey}
and updating~\cite{bellini2020blockchain} reputation scores could also
incorporate factors such as the number of IoT requests handled or compliance
with service level agreements (SLAs)~\cite{govindaraj2021review}.
The collateral requirement $D$ may decrease for fog nodes with sustained
positive audit histories~\cite{kozhan2021decentralized}, allowing excess
collateral to be withdrawn. 

\paragraph{Adaptive Audit Rates and Penalties}

Audit frequency and penalty severity may be adjusted dynamically to accelerate the removal of non-compliant fog nodes. For example, audit frequency may increase for nodes with poor verification histories~\cite{lewis1979simulation}, while penalties may escalate following repeated violations~\cite{miceli2005simple}.

\paragraph{Penalty Delay}

A penalty issued shortly after a service transaction may allow a fog node to link the transaction to a covert audit and infer the IoT identity used by the oracle. Therefore, penalty enforcement may be delayed or aggregated across multiple audit cycles to strengthen oracle anonymity~\cite{javidbakht2017delay}.

\medskip

Overall, these mechanisms extend the baseline design while improving system flexibility, robustness, and fairness. Their integration into FISIE is outside the scope of this work.

\section{Conclusion} \label{sec-conclusion}

The integrity of fog nodes in IoT environments remains a critical challenge. This paper introduces the blockchain-based FISIE framework for decentralized IoT--fog environments. FISIE enables mutual authentication, streamlines service transactions, and enforces integrity through audit-driven rewards and penalties.

We formulated a latent-state dynamic graphical model to describe how fog-node behaviour influences service traffic and profit across successive audit cycles. Discrete-event simulation shows that behavioural strategies affect short-term outcomes through their impact on reputation and net profit. However, long-term survival is governed by audit-driven integrity enforcement, since repeated verification failures ultimately lead to ejection. Therefore, only full honesty sustains long-term participation and profitability.

Future work will evaluate the real-world feasibility of blockchain-enabled integrity enforcement in fog environments. A proof-of-concept implementation will assess network latency, packet loss, energy consumption, and fog-node behaviour under realistic deployment conditions.

This work demonstrates that penalty enforcement through blockchain-based auditing provides effective incentives for enforcing integrity in decentralized IoT--fog systems. Ensuring fog integrity will help maintain a stable and secure execution environment, enabling the expansion of IoT applications toward real-world smart cities~\cite{zhang2020design}.

\section*{Acknowledgement}
We acknowledge the support of the Natural Sciences and Engineering Research Council of Canada (NSERC), [CGS D - 558695 - 2021].

\bibliographystyle{ieeetr}
\bibliography{references}
\end{document}